\documentclass[twocolumn,secnumarabic,amssymb, nofootinbib,aps, superscriptaddress, pra]{revtex4-1}

\usepackage{amsthm}
\usepackage{amsmath}
\usepackage{amssymb}
\usepackage{color}
\usepackage{adjustbox}
\usepackage{graphicx}
\usepackage{hyperref}

\newcommand{\dd}{\mathrm{d}}

\begin{document}
\title{Quantum nonlocality with arbitrary limited detection efficiency}
\author{Gilles P\"utz}
\email[]{Gilles.Puetz@unige.ch}
\affiliation{University of Geneva}
\author{Nicolas Gisin}
\affiliation{University of Geneva}
\date{\today}

\begin{abstract}
The demonstration and use of nonlocality, as defined by Bell's theorem, rely strongly on dealing with non-detection events due to losses and detector inefficiencies. Otherwise, the so-called detection loophole could be exploited. The only way to avoid this is to have detection efficiencies that are above a certain threshold. We introduce the intermediate assumption of limited detection efficiency, e.g. in each run of the experiment the overall detection efficiency is lower bounded by $\eta_{min}>0$. Hence, in an adversarial scenario, the adversaries have arbitrary large but not full control over the inefficiencies. We analyse the set of possible correlations that fulfil Limited Detection Locality (LDL) and show that they necessarily satisfy some linear Bell-like inequalities. We prove that quantum theory predicts violation of one of these inequalities for all $\eta_{min}>0$. Hence, nonlocality can be demonstrated with arbitrarily small limited detection efficiencies. Finally we propose a generalized scheme that uses this characterization to deal with detection inefficiencies, which interpolates between the two usual schemes, postselection and outcome assignment.
\end{abstract}

\maketitle

\textit{Introduction ---} When studying the discoveries in fundamental physics of the past century one cannot help but come across Bell's seminal work~\cite{Bell1964} on the nonlocal nature of quantum theory. It implies that quantum mechanics can produce correlations which cannot be explained by a common past with local variables propagating contiguously. This has not only proven fascinating from a foundational point of view, but also given rise to applications in device independent quantum information processing~\cite{Brunner:RMP} (DIQIP), like quantum key distribution~\cite{Ekert91,BarrettKent05,Acin06}, randomness generation\cite{Colbeck2006,Pironio2010} or entanglement certification~\cite{Bancal11,Barreiro13}.

\begin{figure}
\includegraphics[width=0.4\textwidth]{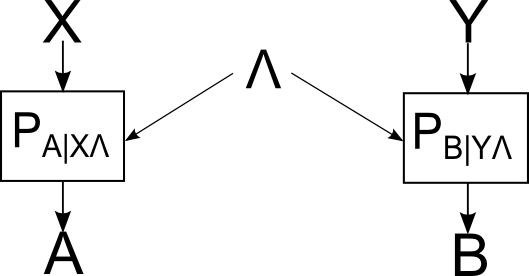}
\caption{Two boxes are programmed by a hidden common strategy $\Lambda$. The boxes are given inputs $X$ and $Y$ and return outputs $A$ and $B$. There is the possibility for nondetection events, in which case the corresponding output variable takes the value $\varnothing$.}
\label{localscenario}
\end{figure}

Let us briefly recall the concept of local and nonlocal correlations. Assume that a source emits particle pairs that travel to two distant labs, in which two experimenters, traditionally called Alice and Bob, perform measurements on them (cf. fig. \ref{localscenario}). Alice locally performs one of several possible measurements and records the outcome, as does Bob. We denote Alice's and Bob's measurement choice by $X$ and $Y$ and their recorded outcomes by $A$ and $B$, respectively\footnote{Notation: we use capital letters to denote random variables and lower case letters to denote the values these variables can take.}. By doing so, they can compute the correlation $P_{AB|XY}$. Given the setup, it seems natural to think that any correlations that Alice and Bob can observe in this way are due to the particles having a common past, as they come from the same source. We refer to this common past by $\Lambda$. Correlations that can be explained by the existence of such a $\Lambda$ are called \textit{local}:
\begin{align}
\label{locality}
P_{L}(ab|xy)=\int\dd\lambda\rho(\lambda)P(a|x\lambda)P(b|y\lambda).
\end{align}
Bell's work showed that there are quantum correlations that cannot be reproduced by such a local model, proving that quantum mechanics is inherently \textit{nonlocal}. This fact has since been demonstrated in a multitude of experiments (see e.g. \cite{Brunner:RMP} and found use in applications~\cite{Ekert91,Acin06,Colbeck2006,Pironio2010,Bancal11}.

However, when demonstrating quantum nonlocality, several issues have to be dealt with. Here we are specifically interested in one of them: what happens if the particles can be lost on the way to or inside the labs, including the possibility that the particles reach the detectors but are simply not registered by them. In this case we say that $A=\varnothing$ or $B=\varnothing$. One immediate idea is of course to carefully analyse why the particles get lost and, if the mechanisms are well understood, to simply discard these cases. This means that Alice and Bob postselect on the cases in which they both registered a detection: $A\neq\varnothing$ and $B\neq\varnothing$. However, this opens up the possibility that fully local correlations appear nonlocal if our understanding of the cause of the non-detections is incorrect, a situation that we wish to avoid~\cite{Branciard11,Pomarico11}. This is especially relevant in the case of active adversaries in DIQIP applications. Another option is to consider the nondetection events as an additional possible outcome and simply check if the resulting correlation is nonlocal. In this case one will never mistake a local correlation for a nonlocal one. The drawback however is that even highly nonlocal distributions may now appear local.

In the end, the only way to deal with this issue, usually called the detection loophole, consists of not only producing highly nonlocal correlations but also having a high enough detection efficiency. If the latter is not satisfied, even a perfect state preparation and perfectly calibrated measurement apparatuses do not help and one is left with an inconclusive experiment unless one assumes that the detection loophole is not exploited.

In this paper, we introduce the concept of limited detection locality. It consists of an intermediate assumption between neglecting the detection loophole and closing it completely. We show that this assumption, even when arbitrarily weak, allows one to demonstrate nonlocality by postselection even with arbitrarily low overall detection efficiency. In addition, we show that the two previously mentioned methods of dealing with detection inefficiencies (postselection and assignment to an additional outcome) can be seen as a special case of a more general method that we present below.

\textit{Limited Detection Locality (LDL)--- } We now introduce the assumption of limited detection efficiency. Assume that there exist a fixed $\eta_{min}$ and $\eta_{max}$ with $[\eta_{min},\eta_{max}]\subsetneqq [0,1]$ such that
\begin{align}
\label{lde}
\eta_{min}\leq P(A\neq\varnothing|x\lambda)\leq\eta_{max}
\end{align}
and similarly for Bob. This corresponds to the assumption that, for any input $x$ and any common local variable $\lambda$, there is a probability of at least $\eta_{min}$ and at most $\eta_{max}$ of having a detection. Consider for example a world in which the polarization degree of freedom of photons was as of yet undiscovered. It is nowadays well known that the detection efficiency of almost all types of detectors is indeed susceptible to polarization. However the detection efficiency never goes up to $1$ or down to $0$, which corresponds to our assumption of limited detection efficiency with nontrivial $\eta_{min}$ and $\eta_{max}$. 
We refer to correlations fulfilling conditions (\ref{locality}) and (\ref{lde}) as \textit{limited detection local}. Note that technically the case of $[\eta_{min},\eta_{max}]=[0,1]$ can still be analysed by our techniques and in fact we would recover the results of Branciard~\cite{Branciard11}.

In an experiment, one can additionally determine the actual observed detection efficiencies, which may of course be different for the different sets of inputs. It is reasonable to assume that some detections occurred for all possible sets of inputs.
\begin{align}
\label{de}
P(a\neq\varnothing|x)=\eta^A_{x}>0.
\end{align}
Since
\begin{align}
P(a\neq\varnothing|x)=\int\dd\lambda\rho(\lambda)P(a\neq\varnothing|x\lambda),
\end{align}
we have that $\eta_{min}\leq\eta^A_x\leq\eta_{max}$. All of this holds analogously for Bob's side. To ease notation, we are going to define $\eta_{xy}=\eta^A_{x}\eta^B_y$.

We can now focus on the postselected limited detection local distributions given by
\begin{align}
P(ab|xy,a\neq\varnothing,b\neq\varnothing)=\frac{P(ab|xy)}{\eta_{xy}}.
\end{align}

Similarly to local correlations, these postselected limited detection local correlations fulfil certain conditions. More precisely, they form a convex polytope~\cite{SuppMat} and therefore respect a set of linear Bell-like inequalities. Making the additional assumption that $\eta_{xy}=\eta_{x'y'}$ for all $x,x',y,y'$, one of these inequalities is, for example, given by
\begin{align}
\label{ldlineq}
\eta_{min}^2P(00|00,a\neq\varnothing,b\neq\varnothing)\nonumber\\
 - \eta_{min}\eta_{max}P(01|01,a\neq\varnothing,b\neq\varnothing)\nonumber\\
-\eta_{min}\eta_{max}P(10|10,a\neq\varnothing,b\neq\varnothing)\nonumber\\
-\eta_{max}^2P(00|11,a\neq\varnothing,b\neq\varnothing)\leq 0.
\end{align}
In a experiment with given losses, the experimenter can check for which values of $\eta_{min}$ and $\eta_{max}$ his observed correlations violate this inequality. He can then conclude that no limited detection local model with these parameters could have reproduced them.

Interestingly, there are quantum correlations that do not fulfil this inequality independent of the observed detection efficiences $\eta_{xy}$ (including the case where the detection efficiency is different for each input pair) and for any upper bound $\eta_{max}$ as long as $\eta_{min}>0$. In fact this is achieved by all quantum correlations violating Hardy's paradox~\cite{Hardy93} and can therefore be realised using any sufficiently pure partially entangled 2-qubit state with the right set of projective measurements. This may be quite surprising since it is well-known that without the assumption of limited detection efficiency (\ref{lde}), a minimal observed detection efficiency of $\eta=\sum_{xy}\eta_{xy}/4>\frac{2}{3}$ is required to demonstrate nonlocality for 2 parties using binary inputs and outputs. However, making the arbitrarily weak additional assumption that $P(a\neq\varnothing|x\lambda)\geq\eta_{min}>0$ allows one to demonstrate quantum nonlocality despite arbitrarily large losses and detection inefficiencies.

\textit{A more general method of dealing with detection inefficiencies--- } It is possible to impose any desired $\eta_{min}$ at the price of adding some noise to the system. Assume that Alice and Bob set their detection systems, which we assume to have an efficiency $\eta$, such that any time a nondetection event occurs, the system still gives an outcome with probability $\eta_{min}$. In this way, Alice and Bob impose that their detection systems have limited detection efficiency given by the chosen $\eta_{min}$ and $\eta_{max}=1$ and they can treat the resulting correlations by the tools presented above. This however comes at the price of adding local noise to their correlations. In fact, assume that Alice and Bob would share the nonlocal correlation $P_{NL}$ if the detectors were perfect and there were no losses (e.g. given by projective measurements on a pure quantum state) and denote by $P_{NL}^A$ and $P_{NL}^B$ the marginal distributions of Alice and Bob respectively. In the nondetection cases the detection systems are set up such that they give with probability $\eta_{min}$ an outcome given by the local distributions $P_L^A$ and $P_L^B$ respectively. Then, by postselecting on the cases where the detection systems gave an outcome, Alice and Bob share the correlation
\begin{align}
P=\Big(\eta^2P_{NL}+\eta(1-\eta)\eta_{min}(P_{NL}^AP_L^B+P_{L}^AP_{NL}^B)\\
+(1-\eta)^2\eta_{min}^2P_L^AP_L^B\Big)\frac{1}{(\eta+(1-\eta)\eta_{min})^2}.
\end{align}
They can then analyse this correlation using the tools of limited detection efficiency presented above.

In fact, in the introduction we mentioned the possibility of dealing with losses and detector inefficiencies by assigning the nondetection events to an additional outcome and treat the resulting correlations using the usual tools of nonlocality. This corresponds exactly to the strategy we just presented with $\eta_{min}=1$. However, our approach is more general, allowing to assign only a fraction of the nondetection events to an outcome and postselecting on the rest. For a fixed detection efficiency $\eta$, our method therefore encompasses both of the previous strategies, full postselection and full assignment to an additional outcome, and additionally allows for an arbitrary mixture of the two. It is at this point not obvious to us that for a given experiment (meaning for a given $P_{NL}$ and a given $\eta$), all of these strategies would yield the same result. We leave it up for future works to analyse this question in more detail.

\textit{Link to measurement dependent locality (MDL)--- } Another way to counterfit nonlocal correlations using only local resources is if the common history $\Lambda$ is correlated with the inputs $X$ and $Y$. If the correlation can be arbitrary, then any nonlocal correlation can be counterfitted in this way, so limitations have to be imposed to be able to make any conclusions. Together with some coauthors, we recently studied the case of measurement dependent local correlations~\cite{putz14} that are defined in the following way:
\begin{align}
\label{MDLcorr}
P(abxy)=\int\dd\lambda\rho(\lambda)P(xy|\lambda)P(a|x\lambda)P(b|y\lambda)\\
\label{MDLcond}
\ell\leq P(xy|\lambda)\leq h.
\end{align}
Note that if Alice and Bob each have $N$ inputs, then $0\leq\ell\leq\frac{1}{N^2}\leq h\leq 1$ due to the normalization of probability distributions. Similarly to this paper, we showed that the set of MDL-correlations for fixed $\ell$ and $h$ can be analysed using Bell-like inequalities. 

It turns out that there exists a strong link between the concepts of limited detection locality and measurement dependent locality. Indeed we make this connection explicit by the following theorem, which we state loosely here and more explicitly in the appendix:

\textit{Theorem: } Assume that we have a correlation that can be produced by using a combination of postselected limited detection (\ref{lde}) and measurement dependent (\ref{MDLcond}) local (\ref{MDLcorr}) resources, with parameters $(\eta_{min},\eta_{max})$ and $(\ell,h)$, respectively. Then this correlation can also be reproduced using only measurement dependent local resources with $\ell'=\frac{\eta_{min}^2}{\eta_{max}^2}\ell$ and $h'=\frac{\eta_{max}^2}{\eta_{min}^2}h$.

Intuitively, the link comes from the fact that the way to exploit postselection for an adversary is to not answer when they do not like the input, resulting effectively, via postselection, in them influencing the inputs. A consequence of this theorem is that whenever a correlation cannot be reproduced by a measurement dependent local model with bounds $\ell'$ and $h'$, then it can also not be realized using limited detection efficiencies with $\frac{\eta_{min}^2}{\eta_{max}^2}\geq N^2\ell$ and $\frac{\eta_{max}^2}{\eta_{min}^2}\leq N^2 h$ where $N$ is the number of inputs for each of the two parties. This allows us to use any result derived for the MDL-scenario and apply them to LDL-correlations. Even more interestingly, we are now able to deal with the problems of losses and measurement dependence in a straightforward way since we can simply focus exclusively on measurement dependence.

\textit{Conclusion --- } Losses and detection inefficiencies have been a long-lasting thorn in all experimenters side. They are a big part of the reason that a loop-hole free Bell-test has to this day not been conducted while also being one of the main weak points that an adversary will attack in any task whose security relies on quantum mechanics. To help deal with both of these issues from a theoretical point of view, we introduced the additional assumption of limited detection efficiency (\ref{lde}). The assumption at its core corresponds to assuming that the inefficiencies in the setup are only partially exploited, an idea that we consider very intuitive in its nature. For the case of a fundamental Bell test, assuming that nature is non-malicious, this idea seems very natural. However, when dealing with an adversary in device independent quantum information processing tasks, it is less obvious to motivate the assumption in the general scenario due to detector blinding attacks and similar measures. The concept can be used to draw stronger conclusions in any experiment that does not fully close the detection loophole. Its main appeal lies in the fact that even when the limitation assumption is made arbitrarily weak, the nonlocal nature of quantum mechanics can still be revealed with arbitrarily low overall detection efficiency.

In addition, we introduced a generalized method to deal with detection inefficiencies in general. The method includes the usual methods of postselection and assignment to an additional outcome as special cases and allows to mix the two. It is an open question whether or not for a given experiment one of this continuum of methods trumps the other ones or if they are all equaivalent. We leave this question for future research.

Finally, we connected the ideas of limited detection locality and measurement dependent locality. We showed that in fact results from studying measurement dependent local correlations can be applied to the case of limited detection locality. Moreover, it is possible to deal with detection efficiencies and lack of measurement independence at the same time, which we hope will be of use for future Bell experiments.

\textit{Acknowledgments --- } We acknowledge Alex May who wrote down an early version of a proof similar to the one given in appendix 2 showing a connection between limited detection and measurement dependence, as well as Roger Colbeck, who mentioned the idea of limited detection in private discussions. We further acknowledge financial support by the European project CHIST-ERA DIQIP as well as the Swiss NCCR-QSIT.

\bibliography{LDLpaper2}

\newpage
\onecolumngrid
\appendix

\section{Polytopal structure of Limited Detection Local correlations}
Consider the case where $N$ parties perform a nonlocality experiment. The input and outcome of the $i$-th party will be denoted by $X_i$ and $A_i$ respectively. We consider the case where nondetection events can occur, they will be denoted by $A_i$ taking the value $\varnothing$. We will denote by $A'_i$ the outcome of party $i$ after postselecting on having a detection. As discussed in the maintext, we make the assumption of limited detection locality and show that these correlations form a polytope. The theorem stated here is more general than needed for the maintext, where we only consider the case of 2 parties.\\

\textit{Definitions: } Let $\{A_i\}_{i=1}^N$, $\{A'_i\}_{i=1}^N$, $\{X_i\}_{i=1}^N$ be sets of random variables with alphabets $\{1\cdots m_i,\varnothing\}$, $\{1\cdots m_i\}$ and $\{1\cdots n_i\}$ respectively, $m_i,n_i\in\mathcal{N}$. In the following, the corresponding lower case letters will denote values in the respective alphabet. We will denote probability distributions over a random variable $V$ by $P_V$, the value of this distribution for a given value of $V$ by $P_V(v)$. For ease of notation, we will often omit the random variable and just write $P(v)$. We will denote conditional probability distributions over a random variable $V$ conditioned on a random variable $W$ by $P_{V|W}$. In the case of continuous random variable we denote the probability density by $\rho_V$. In the following we assume that all the probability distributions are well defined. The set of all probability distributions over $V$ will be denoted by $\mathcal{P}_V$ and of all conditional probability distributions over $V$ conditioned on $W$ by $\mathcal{P}_{V|W}$.\\

We define the following sets:

\begin{itemize}
\item The sets of 1-party distributions with limited detection:
\begin{align*}
\mathcal{LD}_i(\eta_{min},\eta_{max})=\Big\{ P_{A_i|X_i}\in\mathcal{P}_{A_i|X_i} : 1-\eta_{min}\leq P_{A_i|X_i}(\varnothing|x)\leq 1-\eta_{max} \forall x\in\{0\cdots n_i\} \Big\}
\end{align*}
\item The set of $N$-party limited detection local distributions:
\begin{align*}
\mathcal{LDL}_N(\{\eta_{min,i}\}_{i=1}^N,\{\eta_{max,i}\}_{i=1}^N) =\Big\{P_{A_1\ldots A_N|X_1\ldots X_N}\in&\mathcal{P}_{A_1\ldots A_N|X_1\ldots X_N} : \exists \Lambda \text{ s.t. } \\
&P(a_1\ldots a_N|x_1\ldots x_N) = \int\dd\lambda\rho(\lambda)\prod_{i=1}^N P(a_i|x_i\lambda), \\
&P_{A_i|X_i \Lambda=\lambda}\in \mathcal{LD}_i(\eta_{min,i},\eta_{max,i}) \forall\lambda, \forall i\Big\}
\end{align*}
\item The set of $N$-party postselected limited detection local distributions:
\begin{align*}
\mathcal{LDLPS}_N(\{\eta_{min,i}\}_{i=1}^N,\{\eta_{max,i}\}_{i=1}^N,\{\eta_{x_1\ldots x_N}\}_{x_1\ldots x_N})=\Big\{&P_{A'_1\ldots A'_N|X_1\ldots X_N}\in\mathcal{P}_{A'_1\ldots A'_N|X_1\ldots X_N} : \\
&\exists Q_{A_1\ldots A_N|X_1\ldots X_N}\in\mathcal{LDL}(\{\eta_{min,i}\}_{i=1}^N,\{\eta_{max,i}\}_{i=1}^N), \\
&Q(a_1\neq\varnothing\ldots a_N\neq\varnothing|x_1\ldots x_N)=\eta_{x_1\ldots x_N},\\
&P(a'_1\ldots a'_N|x_1\ldots x_N) = \frac{Q(a'_1\ldots a'_N|x_1\ldots x_N)}{\eta_{x_1\ldots x_N}}\Big\}.
\end{align*}
\end{itemize}

We further define the following two sets, which we will prove to be the vertices of $\mathcal{LD}_i(\eta_{min},\eta_{max})$ and $\mathcal{LDL}_N(\{\eta_{min,i}\}_{i=1}^N,\{\eta_{max,i}\}_{i=1}^N)$:
\begin{align*}
\mathcal{V}^{\mathcal{LD}_{i}}(\eta_{min},\eta_{max})=\Big\{ V_{A_i|X_i}\in&\mathcal{P}_{A_i|X_i} : \forall x\in\{1\ldots n_i\}\exists! a_x\in\{1\ldots m_i\} \text{ and }\eta_x\in\{\eta_{min},\eta_{max}\} \text{ s.t.}\\
&V(a_x|x)=\eta_x \text{, } V(\varnothing|x)=1-\eta_x \text{ and otherwise } V(a|x)=0 \Big\}
\end{align*}
\begin{align*}
\mathcal{V}^{\mathcal{LDL}_N}(\{\eta_{min,i}\}_{i=1}^N,\{\eta_{max,i}\}_{i=1}^N) =\Big\{V_{A_1\ldots A_N|X_1\ldots X_N}\in&\mathcal{P}_{A_1\ldots A_N|X_1\ldots X_N} : \\
&\exists V_i\in\mathcal{V}^{\mathcal{LD}_{i}}(\eta_{min,i},\eta_{max,i}) \text{ s.t. }\\
&V(a_1\ldots a_N|x_1\ldots x_N) = \prod_{i=1}^NV_i(a_i|x_i)\Big\}
\end{align*}

With these definitions, we can now state the theorem. It refers to polytopes, which for the purposes of this work are simply seen as a convex structure with a finite set of vertices. Equivalently they can be defined by a finite set of inequalities.\\

\textbf{Theorem: } For fixed $N$, $\{\eta_{min,i}\}_{i=1}^N$ and $\{\eta_{max,i}\}_{i=1}^N$, $\mathcal{LDL}_N(\{\eta_{min,i}\}_{i=1}^N,\{\eta_{max,i}\}_{i=1}^N)$ is a polytope whose vertices are a subset of $\mathcal{V}^{\mathcal{LDL}_N}(\{\eta_{min,i}\}_{i=1}^N,\{\eta_{max,i}\}_{i=1}^N)$. Furthermore, for fixed $\{\eta_{x_1\ldots x_N}\}_{x_1\ldots x_N})$, $\mathcal{LDLPS}_N(\{\eta_{min,i}\}_{i=1}^N,\{\eta_{max,i}\}_{i=1}^N,\{\eta_{x_1\ldots x_N}\}_{x_1\ldots x_N})$ is also a polytope.\\

\textit{Proof: }To ease notation we will omit writing $N$, $\{\eta_{min,i}\}_{i=1}^N$ and $\{\eta_{max,i}\}_{i=1}^N$ from now on. The first part of the theorem follows from the following two lemmas.\\

\textbf{Lemma 1: } $\mathcal{LD}_i$ is a polytope with vertices given by $\mathcal{V}^{\mathcal{LD}_{i}}$.\\

\textit{Proof: } Due to the normalisation of probability distributions, i.e.
\begin{align*}
\sum_{a_i}P(a_i|x_i)=1 \text{ }\forall x_i,
\end{align*}
we have that $P(\varnothing|x_i)=1-\sum_{a_i=1}^{m_i}P(a_i|x_i)$ and we can therefore work in the lowerdimensional subspace given by $a_i\in\{1\ldots m_i\}$. In this subspace, we are then left with the polytope defined by the inequalities $\eta_{min}\leq\sum_{a_i=1}^{m_i}P(a_i|x_i)\leq\eta_{max}$. This is the definition of a hypercube whose vertices are defined by the corresponding part of $\mathcal{V}^{\mathcal{LD}_i}$.\\

The Lemma follows. $\square$\\

\textbf{Lemma 2: } Let $\mathcal{Q}$ and $\mathcal{R}$ be polytopes with vertices $\mathcal{V_Q}$ and $\mathcal{V_R}$ respectively. Let $\mathcal{S}=\Big\{S: \exists\Lambda\text{ s.t. } S(u,v)=\int\dd\lambda\rho(\lambda)Q_\lambda(u)R_\lambda(v) \text{ with }Q\in\mathcal{Q},R\in\mathcal{R}\Big\}$. Let $\mathcal{V_S}=\Big\{V_S : V_S(u,v)=V_Q(u)V_R(v) \text{ with }V_Q\in\mathcal{V_Q}, V_R\in\mathcal{V_R}\Big\}$.

Then $\mathcal{S}$ is a polytope whose vertices are a subset of $\mathcal{V_S}$.\\

\textit{Proof: } By definition, $\mathcal{S}$ is convex and $\mathcal{V_S}\in\mathcal{S}$.

Let $S\in\mathcal{S}$, then by definition we have:
\begin{align*}
S(u,v)&=\int\dd\lambda\rho(\lambda)Q_\lambda(u)R_\lambda(v)\\
&=\int\dd\lambda\rho(\lambda)\sum_{i}q_{\lambda,i}V_Q^{i}(u)\sum_{j}r_{\lambda,j}V_R^{j}(v)\\
&=\sum_{(ij)}\Big(\int\dd\lambda\rho(\lambda)q_{\lambda,i}r_{\lambda,j}\Big)V_Q^{i}(u)V_R^{j}(v)\\
&=\sum_{(ij)}s_{ij}V_S^{ij}(u,v)
\end{align*}
In the first step we use the fact that any element of $\mathcal{Q}$ and $\mathcal{R}$ can be written as a convex combination of their vertices and we define $q_{\lambda,i}\geq 0$, $\sum_i q_{\lambda,i}=1$ and $r_{\lambda,j}\geq 0$, $\sum_j r_{\lambda,j}=1$. In the last step we defined $s_{ij}=\int\dd\lambda\rho(\lambda)q_{\lambda,i}r_{\lambda,j}$, which fulfils $s_{ij}\geq 0$ and $\sum_{ij}s_{ij}=1$ and also used the definition of $V_S^{ij}$.\\

This proves the Lemma.$\square$\\

Using these two Lemmas in conjunction (and using Lemma 2 iteratively) proves that $\mathcal{LDL}$ is a polytope and that $\mathcal{V_LDL}$ contains its vertices.

To finalize the proof of the theorem we need to show that $\mathcal{LDLPS}$ is a polytope as well. This can be seen directly since the set is obtained by slicing $\mathcal{LDL}$ with the hyperplanes defined by $P(a_1\neq\varnothing\ldots a_N\neq\varnothing|x_1\ldots x_N)=\eta_{x_1\ldots x_N}$. Cutting a polytope with hyperplanes results in another polytope. The final step is a simple rescaling of the entries (equivalent to rescaling the axes) and therefore the set remains again a polytope. \\

This proves the theorem. $\square$\\

\section{Limited Detection Locality and Measurement Dependent Locality}
In this section we prove the link between limited detection local and measurement dependent local distributions. This can be proven more generally, here we only present the 2-party version.\\

\textit{Definitions: }We introduce the random variables $D_A$ and $D_B$ with alphabet $\{0,1\}$ such that $D_A=0$ if and only if $A=\varnothing$. We define the set of limited detection local distributions allowing for measurement dependence:
\begin{align*}
\mathcal{MDLDL}(\ell,h,\eta_{min},\eta_{max})=\big\{P_{AD_ABD_BXY} : &P(ad_Abd_Bxy)=\int\dd\lambda\rho(\lambda)P(xy|\lambda)P(ad_Abd_B|xy\lambda), \\
 & \eta_{min}\leq P_{D_AD_B|XY\Lambda}(11|xy\lambda)\leq\eta_{max},\\ 
 &P(ad_Abd_B|xy\lambda)=P(ad_A|x\lambda)P(bd_B|y\lambda)\\
&\ell\leq P(xy|\lambda)\leq h,\\
 &\int\dd\lambda\rho(\lambda)=1, \rho(\lambda)\geq 0\big\}
\end{align*}

We also define the set of measurement dependent local correlations:
\begin{align*}
\mathcal{MDL}(h,\ell)=\big\{P_{ABXY} : &P(abxy)=\int\dd\lambda\rho(\lambda)P(xy|\lambda)P(ab|xy\lambda), \\
 &P(ab|xy\lambda)=P(a|x\lambda)P(b|y\lambda)\\
&\ell\leq P(xy|\lambda)\leq h,\\
 &\int\dd\lambda\rho(\lambda)=1, \rho(\lambda)\geq 0\big\}
\end{align*}


\textit{Theorem: } If 
\begin{align*}
P_{AD_ABD_BXY}&\in \mathcal{MDLDL}(\ell,h,\eta_{min},\eta_{max})
\end{align*}
then
\begin{align*}
P_{ABXY|D_A=1,D_B=1}\in \mathcal{MDL}(\frac{\eta_{min}}{\eta_{max}}\ell,\frac{\eta_{max}}{\eta_{min}}h).
\end{align*}

\textit{Proof: }
We have
\begin{enumerate}
\item $P(ad_Abd_Bxy)=\int\dd\lambda\rho(\lambda)P(xy|\lambda)P(ad_Abd_B|xy\lambda)$
\item $P(ad_Abd_B|xy\lambda)=P(ad_A|x\lambda)P(bd_B|y\lambda)$
\item $\eta_{min}\leq P_{D_AD_B|XY\Lambda}(11|xy\lambda)\leq\eta_{max}$
\item $\ell\leq P(xy|\lambda)\leq h$.
\end{enumerate}

Let us prove a few implications:
\begin{itemize}
\item If $(AD_A|X)$ and $(BD_B|Y)$ are local, meaning that they fulfil condition 2, then $(D_A|X)$ and $(D_B|Y)$ are also local:
\begin{align*}
P(d_Ad_B|xy\lambda)&=\sum_{a,b}P(ad_Abd_B|xy\lambda)\\
&=\sum_{ab}P(ad_A|x\lambda)P(bd_B|y\lambda)\\
&=P(d_A|x\lambda)P(d_B|y\lambda).
\end{align*}
\item If $(AD_A|X)$ and $(BD_B|Y)$ are local, then $(A|D_AX)$ and $(B|D_BY)$ are also local:
\begin{align*}
P(ab|xd_Ayd_B\lambda)&=\frac{P(ad_Abd_B|xy\lambda)}{P(d_Ad_B|xy\lambda)}\\
&=\frac{P(ad_A|x\lambda)}{P(d_A|x\lambda)}\frac{P(bd_B|y\lambda)}{P(d_B|y\lambda)}\\
&=P(a|xd_A\lambda)P(b|yd_B\lambda).
\end{align*}
\item Knowing less cannot result in knowing more, meaning that upper and lower bounds on $P(\mu|\nu\sigma)$ also hold for $P(\mu|\nu)$: Assume $P(\mu|\nu\sigma)\leq h$, then
\begin{align*}
P(\mu|\nu)&=\sum_{\sigma}P(\sigma)P(\mu|\nu\sigma)\\
&\leq h\sum_\sigma P(\sigma)\\
&=h
\end{align*}
where we used that $\sum_\sigma P(\sigma)=1$. The same holds for lower bounds $\ell\leq P(\mu|\nu\sigma)$. Due to this, condition 3 implies 
\begin{align*}
\eta_{min}\leq P_{D_AD_B|\Lambda}(11|\lambda)\leq\eta_{max}.
\end{align*}
\end{itemize}
Using the implications above, we can show that the conditions imply bounds on $P(xy|D_A=1,D_B=1,\lambda)$:
\begin{align*}
P(xy|D_A=1,D_B=1,\lambda)=\frac{\overbrace{P(D_A=1,D_B=1|xy\lambda)}^{\eta_{min}\leq\cdots\leq\eta_{max}}}{\underbrace{P(D_A=1,D_B=1|\lambda)}_{\eta_{min}\leq\cdots\leq\eta_{max}}}\underbrace{P(xy|\lambda)}_{\ell\leq\cdots\leq h}\\
\Rightarrow \frac{\eta_{min}}{\eta_{max}}\ell\leq P(xy|D_A=1,D_B=1,\lambda)\leq\frac{\eta_{max}}{\eta_{min}}h.
\end{align*}

We can now prove the theorem:
\begin{align*}
P(abxy|D_A=1,D_B=1)&=\int\dd\lambda\rho(\lambda|D_A=1,D_B=1)P(xy|\lambda,D_A=1,D_B=1)P(ab|xy\lambda,D_A=1,D_B=1)\\
&= \int\dd\lambda\rho(\lambda|D_A=1,D_B=1)\overbrace{P(xy|\lambda,D_A=1,D_B=1)}^{\frac{\eta_{min}}{\eta_{max}}\ell\leq\cdots \leq\frac{\eta_{max}}{\eta_{min}}}\cdot\\
&\quad \quad\quad P(a|x\lambda,D_A=1)P(b|y\lambda,D_B=1)
\end{align*}
This is by definition an MDL-correlation:
\begin{align*}
P(ABXY|D_A=1,D_B=1)\in\mathcal{MDL}(\frac{\eta_{min}}{\eta_{max}}\ell,\frac{\eta_{max}}{\eta_{min}}h)
\end{align*}

\end{document}